\documentclass{aa501}
\usepackage{graphicx}
\usepackage[english]{babel}

\begin{document}

\title{The Ly-edge paradox and the need for obscured QSOs}

\author{R. Maiolino\inst{1} \and M. Salvati\inst{1}
\and  A. Marconi\inst{1} \and R. R. J. Antonucci\inst{2}}
 
\offprints{R. Maiolino}

\institute{Osservatorio Astrofisico di Arcetri,
  Largo E. Fermi 5, 50125 Firenze, Italy
  \and
Physics Department, University of California,
 Santa Barbara, CA 93106
  }

 \date{Received ; accepted }

\abstract{
Based on the most recent QSO ultraviolet spectra,
the covering
factor of the clouds of the Broad Line Region (BLR)
is about 30\%, or larger. 
This value would imply that in at least 30\% of the QSOs
our line of sight
crosses one, or more, BLR clouds
and, in the latter case, the UV spectrum should
show a sharp Ly-edge in absorption. This Ly-edge in absorption is
never observed. This paradox is solved if, as suggested by various
authors, the BLR is flattened and
the dusty gas  in the outer parts, on the same plane, prevents the
observation along the lines of sight passing through the BLR clouds.
The objects observed edge-on (with respect to the flattened BLR)
 would be classified
as obscured QSOs or, within the framework of the unified model, type~2 QSOs.
The covering factor of the BLR constrains the fraction
of obscured QSOs to be QSO2/QSO1~$>$~0.5. This lower limit is already high
with respect to the number of candidate type 2 QSOs claimed so far.
We discuss this constraint in relation to recent AGN surveys.
}

\maketitle

\section{Introduction}
The existence of obscured QSOs, or type 2 QSOs, has been
a hotly debated issue since the formulation of the first unified theories.
While the relation between Seyfert 1s and their obscured counterparts, the
Seyfert 2s, has been thoroughly assessed, at higher
luminosities ($L_{bol}>10^{45}$erg s$^{-1}$) we are still at the stage
of questioning whether a significant population of
the obscured counterparts of QSOs exist or not.
Only for the sub-class of radio-loud QSOs a population of obscured
counterparts was identified (the narrow line radio galaxies). 
Yet, radio-loud AGNs are only a minor fraction of the whole AGN population.
Among radio-quiet AGNs there are only a few cases of candidate type~2 QSOs.
Most of the ultraluminous infrared galaxies (ULIRGs, which are among the
best candidates to host hidden QSOs) appear mostly powered by starburst
activity, while evidences for a hidden AGN is only found on the high luminosity
tail of this population (Lutz et al. \cite{lutz}, Evans et al. \cite{evans},
Murphy et al.
\cite{murphy}, Veilleux et al. \cite{veilleux}, Genzel et al. \cite{genzel}).
Also, even in those ULIRGs showing evidence
for a hidden AGNs, its contribution to the bolometric luminosity is not
well constrained, except for a few cases (eg. Franceschini et al.
\cite{franceschini}). For what concerns surveys in the hard X-rays, which
should be less affected by obscuration than optical and soft X-ray surveys,
only two cases of type 2 QSO have been identified (Norman et al.
\cite{norman}, Crawford et al. \cite{crawford}).

While the evidence for a ``classical'' population of type 2 QSOs is marginal,
a population of broad-line QSOs whose continuum
is redder than classical color-selected
AGNs was found in radio and X-ray surveys (Webster et al. \cite{webster},
Kim \& Elvis \cite{kim}, Fiore et al. \cite{fiore99},
Risaliti et al. 20001,
Maiolino et al. \cite{maiolino00}).
Although in some of these objects
the red continuum is ascribed to the contribution from the host galaxy
or to synchrotron emission (especially for the radio-loud objects),
there is evidence that in several of them dust reddening and
extinction play a major role.

In this paper we demonstrate that
evidence for the existence of a large population
of optically obscured QSOs is inferred, ironically, by the UV spectra of
unobscured QSOs.

\section{The covering factor of the BLR}

The covering factor of the BLR has been discussed by several
authors in the past, based on the comparison between the hydrogen
recombination lines and the slope of the optical-UV continuum extrapolated
beyond the Ly-edge or by using models for the shape of the
ionizing continuum. By using the Ly$\alpha$ as a photon
counter it was found that the covering factor is between
5\% and 10\% (eg. Oke \& Korycansky \cite{oke},
Carswell \& Ferland \cite{carswell}). By using the EW of H$\beta$ Binette
et al. (\cite{binette}) derived an even higher covering factor ($\sim$40\%).

Recent UV spectra of QSOs have shown that the fraction of ionizing
photons is significantly lower than estimated in the past.
Zheng et al. (\cite{zheng}, hereafter Z97)
have created a composite UV spectrum out of HST
spectra of 101 QSOs with redshift between 0.33 and $\sim$1.5 (i.e.
without large effects from the intervening Ly$\alpha$ forest absorption).
One of the most interesting features of the composite spectrum is that
the slope in the extreme UV region is significantly steeper than was
thought in the past. More recently, Kriss (\cite{kriss})
generated a composite
with 248 QSO spectra observed with HST; the result 
is in good agreement with Z97.
As first noted by Laor et al. (\cite{laor}) 
the fraction of ionizing photons is about a factor of four lower than
in models adopted previously (when normalized to the near-UV/optical
light). Since the equivalent width of the emission lines in Z97
is not very different from previous measurements, this implies that the
covering factor must be higher by a factor of
about four than estimated in the past. More specifically,
by simply using the Ly$\alpha$ flux in the composite spectrum of Z97
as a ionizing photon counter, along
with the number of ionizing photons actually derived from the same spectrum,
a covering factor of 27\% is inferred.

By using more complex models of the BLR the inferred covering factor
does not change significantly. For instance, by using the LOC
model by Korista et al. (\cite{korista98}),
adapted to the spectral shape observed by Z97, and by simply comparing the
expected EW of Ly$\alpha$ and HeII 1640\AA \ with the values measured
by Z97, we obtain a
covering factor of about 30\%.

Several authors have proposed that the ionizing radiation is not
emitted isotropically by the accretion disk and that the edge-on
lines of sight see a harder continuum (eg. Netzer \cite{netzer},
Sun \& Malkan \cite{sun}, Laor \& Netzer \cite{laor}).
If, as proposed by Netzer (\cite{netzer}) and by Korista
et al. (\cite{korista97}) and as discussed more in detail below, 
the disk is preferentially observed pole-on by us,
while the BLR is flattened on the plane of the disk,
then the anisotropic emission implies an even higher covering factor.

A larger covering factor is also obtained if the composite spectrum of Z97
is included in the calculations of Binette et al. (\cite{binette}).

Summarizing, we will assume 30\% as a lower limit to the covering factor
of the BLR.

\begin{figure}
\centering
\includegraphics[width=8truecm]{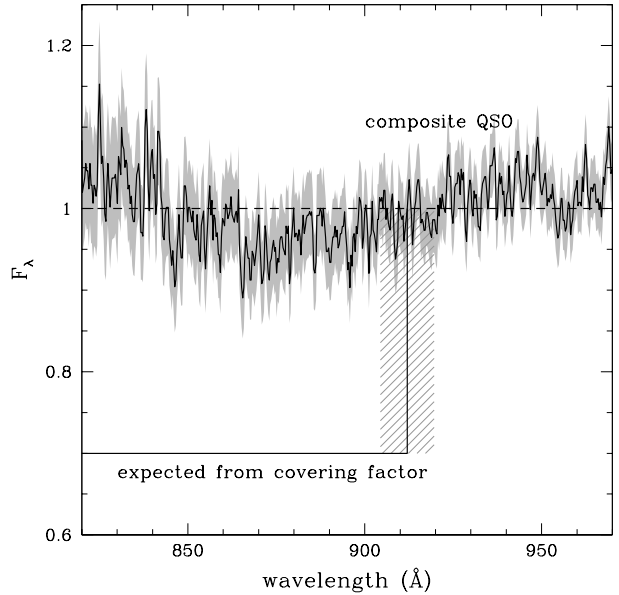}
 \caption{Composite QSO spectrum derived by Z97, divided by the power-law
fitting the continuum in the vicinity of the Ly-edge (and normalized by
the continuum at the Ly-edge). The shaded region gives the error on the
mean. The solid broken line gives the expected depth of the Ly-edge by
assuming a covering factor of 30\% (which is the lower limit on the
derived in sect.2). The hatched region gives the
uncertainty due to the velocity spread of the BLR clouds.}
\end{figure}

\section{The missing Ly-edge}

If the BLR clouds had a spherically symmetric distribution
around the ionizing source then,
given the relatively large covering factor of the BLR,
we would expect that the line of sight towards the
UV ionizing source intercepts a cloud in more than 30\% of the cases,
hence we would expect a sharp Lyman absorption edge in more than
30\% of the QSOs or,
equivalently, the composite spectrum should show a sharp Lyman absorption
edge deeper than 30\% of the continuum red-ward of the Ly-edge.
This remains true even
in the case that the ionizing UV source has dimensions
comparable or larger than the clouds.

Fig.~1 shows the composite QSO spectrum of Z97 (both radio-quiet and
radio-loud QSOs) around the Ly-edge
divided by the power-law which has been fitted to the spectrum in the same
region and normalized at the Ly-edge.
The shaded region indicates the error on the mean.
The composite spectrum does not show
any indication
for the sharp absorption edge expected by absorbing gas in the BLR along our
line of sight. The lower straight line in Fig.~1 shows the upper
limit on the continuum level
blue-ward of the Ly-edge expected from the covering factor of
the BLR, which is clearly inconsistent with the composite QSO spectrum. 

We can even state that, to our knowledge, no single
QSO has shown evidence for the presence of a Ly absorption edge ascribed
to intrinsic absorption from BLR clouds along the line of sight
(Antonucci et al. \cite{antonucci}, Koratkar \& Blaes \cite{koratkar}).
Some sharp Ly-edge absorption in a few QSOs was totally
ascribed to external systems at lower redshift.

This is not the first time that the problem of the missing Ly-edge
is highlighted (eg. Antonucci et al. \cite{antonucci}). Yet, the larger
covering factor of the BLR inferred by the recent data makes this issue
even more serious.

One might argue that if the absorbing
clouds are moving along our line of sight (outflowing and/or inflowing) then
the resulting Ly-edge in the composite spectrum would be broadened and
therefore more difficult to detect. However, given the velocity dispersion
in the BLR, the broadening of the Ly-edge would be at most as high as the
width of the emission lines and, therefore, an
absorption edge as deep as 30\% of the continuum should be detectable.
The hatched vertical region in Fig.~1 shows the width of of the Ly-edge
which would be produced by a velocity dispersion of 5000 km/s.

The emission of a Ly continuum, expected at some level by BLR
models (Korista \& Goad \cite{kg}, Davidson \cite{davidson}),
could partly fill up the Ly edge absorption.
However,
such a Ly continuum emission has never been observed in any object, not
even in those showing a Balmer continuum.
Also, while the photoelectric absorption should extend
at least to the soft X-rays (N$_H$(BLR)$\approx 10^{22}-10^{23}$cm$^{-2}$),
the Ly continuum emission would rapidly decrease blue-ward of the Ly-edge
and at 700\AA\ would be less than 10\% of the value at the Ly-edge.
On the contrary, the Z97 spectrum
does not show any indication of absorption down
to 300\AA.

Binette et al. (1993, and references therein) proposed that the
observed UV-optical continuum consists mostly of 
scattered/reprocessed radiation while the intrinsic spectrum is
obscured along our line of sight in most AGNs (even in type 1). This would
explain the lack of Ly-edge absorption. However, the intrinsic spectrum
should be observable in the hard X-rays with a prominent photoelectric cut-off,
while most of the optically selected QSOs (which roughly
match the sample used by Z97) observed in the X-rays do not show any absorption
cutoff, with only a few exceptions (Reeves \& Turner \cite{reeves},
Maiolino et al. \cite{maiolino01b}).

Dust within the BLR clouds would possibly explain the lack of the Ly-edge
absorption, since it
would absorb the UV continuum also red-ward of the Ly-edge.
However, the BLR is located inside the sublimation radius of dust
and even the shielding the ionizing radiation
due to the gas is probably unable to allow the survival of grains in the
dark side of the clouds.

\section{The paradox and the need for obscured QSOs}

The paradoxical result of this analysis is that although the BLR clouds
cover a large fraction of the sky to the nuclear source, whenever we observe a
QSO our line of sight never intercepts these clouds.

The origin of this paradox is probably related to the assumption that the BLR
has a spherically symmetric distribution around the ionizing source.
If the BLR clouds are mostly distributed in a flattened geometry
and the observer's lines of sight are mostly pole-on, then
the BLR emission is observed, but no Ly-edge in absorption is detected.
Then the edge-on lines of sight, which should intercept the ionized
clouds and show the Ly-edge absorption, are never observed because
of the dusty gas on the plane of the BLR, but outside the
sublimation radius,
and which is
responsible for absorbing and reddening the QSOs observed ``edge-on'',
therefore preventing their detection in color-selected surveys.

Within the framework of the unified model the outer dusty gas
would be identified with the ``obscuring torus'' and the QSOs observed
edge-on would be identified as type 2, obscured QSOs.
The fact that no QSO with Ly absorption edge has ever been observed
implies that
not only the dusty torus must be coplanar with the flattened broad line
region, but also that the covering factor of the torus must be larger
than the covering factor of the broad line region.
Within this scenario a lower limit to the fraction of obscured QSOs is given by
the covering factor of the BLR.
As a consequence, obscured (type 2) QSOs must at least be as numerous as
half of the unobscured population of QSOs.

A flattened distribution for the BLR was proposed by various authors
in the past, both based on theoretical and on observational
arguments (Netzer \cite{netzer}, Collin-Souffrin \& Dumont \cite{collin},
 Wills \& Brotherton \cite{wills}, Wanders et al. \cite{wanders},
 Goad \& Wanders \cite{goad}).
We note that the scenario of a flattened BLR coplanar with the obscuring
torus is consistent with the finding that the hard X-ray spectra of some
Seyfert 2 galaxies show evidence for a double absorber (see review
in Maiolino \cite{maiolino01a}).
While the (total) absorber with lower N$_H$ is ascribed to the obscuring
torus, the absorber with higher N$_H$ (typically a few times
10$^{23}$cm$^{-2}$) has a partial covering (generally $>$30\%) and
it is ascribed to BLR clouds along the
line of sight.

\section{Enough obscured QSOs?}

The problem of the shortage of obscured QSOs is not new and
has been widely discussed in the past. However, with the exception of
the requirement of X-ray absorbed QSOs to account for the X-ray number
counts by Gilli et al. (\cite{gilli}), there were no other observational or
theoretical constraints which required the existence of
large population of (optically) obscured QSOs. In this paper
we link the problem of the missing Ly-edge with the issue of the
shortage of obscured QSOs and constrain the minimum fraction of obscured
QSOs. In this section we compare this lower limit with the fraction
of obscured QSO known so far.

The lower limit on the fraction of type 2 QSOs derived in the former section
(QSO2/QSO1~$>$~0.5) is already
quite high when compared with the handful of (candidate)
type 2 QSOs discovered so far. While in the case of radio loud AGNs the fraction
of narrow line radio galaxies is enough
to account for the Ly-edge paradox (NLRG/RLQSO$\sim$1, Singal \cite{singal}),
for what concerns radio-quiet AGNs, as mentioned in the introduction, there
are only a few {\it bona fide} type 2 QSOs discovered so far.

It should be noted that the samples of QSOs used to study their
extreme UV spectrum are generally biased in favor of UV bright objects
and, therefore, heavily
biased against the so-called red QSOs, which were discussed
in the Introduction. If the properties of most
red QSOs are ascribed to dust extinction, then they could be the population
of QSOs observed
 edge-on which should show Ly-edge in absorption due to BLR clouds,
i.e. the population of "obscured" QSOs.
Is the number of red QSO large enough to account for the missing Ly-edge? 
The fraction of red QSOs in soft X-ray selected and radio selected
samples is below 20\% of the whole broad line QSO population
(Kim \& Elvis \cite{kim} and references therein), i.e. lower than
the minimum fraction of obscured QSOs inferred from the BLR
covering factor ($>$30\%).
However, recently Risaliti et al. (\cite{risaliti01}) have found indications,
among grism-selected QSOs, for a larger fraction of red, absorbed objects
($\sim$50\%).

Hard X-rays are probably the most suited for the search of obscured QSOs. 
The identification of the optical counterparts of the recent deep
and medium-deep Chandra surveys has revealed some powerful (type 1) QSOs
and several narrow line AGNs 
with luminosities in the Seyfert range
(Barger et al. \cite{barger},
Giacconi et al. \cite{giacconi},
Fiore et al. \cite{fiore00}, Hornschemeier et al. \cite{hornschemeier},
Tozzi et al. \cite{tozzi}, Mushotzky et al. \cite{mushotzky},
Brandt et al. \cite{brandt00}, \cite{brandt01},
Crawford et al. \cite{crawford}).
Some obscured AGNs with 2--10~keV luminosity of a few times 10$^{44}$~erg/s
have been
referred to as ``obscured QSOs'', but we consider these objects
just as bright Seyfert 2s. For comparison, NGC6240, whose de-absorbed 2--10~keV
luminosity is $\sim 2\times 10^{44}$~erg/s 
(assuming $H_0=70$, Vignati et al. \cite{vignati}), has an infrared
{\it total} luminosity\footnote{$\rm L_{ir}= L(8-1000\mu m)$ ($\sim L_{bol}$)
as defined in Sanders and Mirabel (\cite{sanders}).}
of only $\rm 7\times 10^{11}
L_{\odot}$ ($H_0=70$) which 
is only a factor of two higher than the total IR luminosity of the local
Seyfert 2 archetype NGC1068.
Actually, to our knowledge the Chandra surveys have so far
identified only two type 2 QSOs with luminosities of
L$_{2-10keV}\approx 10^{45}$~erg~s$^{-1}$ 
(Norman et al. \cite{norman}, Crawford et al. \cite{crawford}).
Possibly, most of the QSO2s have still to be identified among the fainter
counterparts of the Chandra sources, but this seems unlikely since several
Sy2s have already been identified (even at high redshift).
Yet, it should be noted that several of the (type 1)
QSOs identified in the Chandra
surveys have a continuum which is redder than color selected QSOs, therefore
suggesting that these Chandra QSOs are affected by some dust reddening. Some
indication of gas obscuration is also inferred from their X-ray spectral slope
which, for some of them,
is slightly harder than in classical QSOs. The fraction of these
mildly obscured QSOs (either red or with harder X-ray spectra) is
certainly enough to account for the Ly-edge paradox.

The limited sky coverage of
the current Chandra surveys is not optimal for a statistical analysis of the
properties of QSOs since their density in the sky is very low.
The hard X-ray surveys obtained by ASCA and BeppoSAX
(Akiyama et al. \cite{akiyama}, Fiore et al. \cite{fiore99},
Fiore et al. \cite{fiore01})
are certainly shallower, but cover
a much larger area of the sky. However, even in these surveys, despite the
large fraction of (type 1) QSOs identified, no clear cases for
type 2 QSOs were found. Yet, a
significant fraction of mildly obscured QSOs was found:
red QSOs or (type 1) QSOs with indication of X-ray absorption.

Summarizing, although only two QSO2s have been identified in the
hard X-ray surveys so far, a fraction of (mildly) obscured QSOs has
been identified which would be consistent with the lower limit on
the obscured-to-unobscured ratio inferred by the Ly-edge paradox.

Another possibility is that most of the type 2 QSOs are Compton thick
and, therefore, have been missed even by the hard X-ray surveys because
they are too dim in the 2--10 keV band. However, this scenario seems
unlikely at least for the most recent surveys, given
that one of the type 2 QSOs discovered in the Chandra fields
is actually a Compton thick object at high redshift (Norman et al.
\cite{norman}), therefore highlighting
the capability of Chandra to detect even Compton thick type 2 QSOs.
Alternatively, these Compton thick type 2 QSOs have already been detected, but
misidentified as Seyfert 2s (or simply Narrow Line AGNs
with L$_{hard-X} < 10^{44}$erg~s$^{-1}$).
Indeed, in the Compton thick case the intrinsic hard
X-ray luminosity would probably be 2--3 orders of magnitude higher
than inferred by the (reflection-dominated) observed flux.
However, the
ratio between the far-IR and the X-ray cosmic background constrains most
of the absorbed AGNs {\it not} to
be ``reflection dominated''. More specifically it can be shown
that a fraction of Compton
thick, reflection-dominated AGNs contributing more than $\sim$10\% of
the 2-10 keV background would over-produce the COBE background at $\sim
140 \mu$m
(see also Risaliti et al. in prep).

\section{Conclusions}

The UV spectra of QSOs indicate that the covering factor of the clouds
of the Broad Line Region must be larger than $\sim$30\%. This covering
factor would imply that more than 30\% of the lines of sight should
intersect a BLR cloud and show a sharp Ly-edge in absorption. This intrinsic
absorption edge has never been observed in single QSOs spectra nor in a
composite spectrum of 100--200 QSOs. This paradoxical result is
in keeping with previous suggestions that
the BLR is not distributed isotropically around the nuclear
source but flattened; within this scenario
the lines of sight passing through the
BLR clouds are not observed in the UV because of dusty gas associated with
the outer parts of the BLR which obscures the whole nuclear region. The
QSOs observed along the latter lines of sight would be classified as red
QSOs or, if the absorption is large enough, as type 2 QSOs.
Within this scenario the covering factor of the BLR provides a lower limit
to the fraction of obscured QSOs and, more specifically, QSO2/QSO1~$>$~0.5.
This lower limit is already much higher than the fraction observationally
obtained so far.

A significant population of QSO2s might still have to be identified in
current hard X-ray surveys. Alternatively, a possibility is that ``pure''
type 2 QSOs are really sparse and the population
of obscured QSOs inferred from the Ly-edge paradox is accounted for by the
mildly obscured QSOs (the red QSOs and
those type 1 QSOs with some X-ray absorption) found in 
hard X-ray surveys. A minor
fraction of type 2 QSOs might be Compton thick; these
have either been lost even by hard X-ray surveys, or have been
misclassified as Seyfert 2s.

If mildly obscured QSOs found in the hard X-ray surveys (which
have N$_H < 10^{22}-10^{23}$ cm$^{-2}$)
account for the obscured QSOs requested by the
missing Ly-edge, when compared with the N$_H$ distribution of Sy2s (most
of which have N$_H > 10^{23}$ cm$^{-2}$, Risaliti et al. \cite{risaliti99}),
this implies that the average N$_H$ decreases with the luminosity.
This finding is supported by similar results on the radio-loud QSOs
(Singal \cite{singal}, Lawrence \cite{lawrence}, Woltjer \cite{woltjer}).

Finally, our results would imply that most of the hard X-ray background
is produced by heavily obscured
Seyfert 2s (as the ongoing Chandra identification campaigns
seem to confirm) and that mildly obscured QSOs only account for the
2--10 keV number counts at intermediate fluxes ($\sim
10^{-13}$erg~s$^{-1}$~cm$^{-2}$, Gilli et al. \cite{gilli}).

\begin{acknowledgements}
We are grateful to Kirk Korista and Lo Woltjer for useful discussions
and to the anonymous referee for useful comments.
We thank Wei Zheng
for providing the electronic version of the composite QSO
spectrum.
This work was partially supported by the
Italian Ministry for
University and Research (MURST) under grant Cofin00--02--36 and
by the Italian Space Agency (ASI) under grant 1/R/27/00.
\end{acknowledgements}

\end{document}